\documentstyle[aps,epsf]{revtex}
\begin{document}
\draft\twocolumn
\title{Measuring quantum optical Hamiltonians}
\author{G. M. D'Ariano and L. Maccone}
\address{Gruppo di Ottica quantistica, Istituto Nazionale di Fisica
della Materia, Sezione di Pavia, via A. Bassi 6, I-27100 Pavia, Italy}
\maketitle
\begin{abstract}
We show how recent state-reconstruction techniques can be used to
determine the Hamiltonian of an optical device that evolves the
quantum state of radiation. A simple experimental setup is proposed
for measuring the Liouvillian of phase-insensitive devices. The
feasibility of the method with current technology is demonstrated on
the basis of Monte Carlo simulated experiments.
\end{abstract}
\bigskip
\narrowtext

In recent years, the possibility of ``measuring'' the quantum state,
after remaining for a longtime a mere theoretical speculation
\cite{despagnat}, eventually entered the realm of true
experiments. From the first experimental demonstration \cite{smithey},
the so called ``homodyne tomography'' technique advanced to the level
of a quantitative state-reconstruction technique
\cite{dariano2,dariano3}, achieving a high degree of reliability in
experiments \cite{schiller}. This state-reconstruction
method is now ready to be used for concrete applications.  \par What
is the practical use of measuring a quantum state?  Apart from the
availability of a kind of ``universal detector''
\cite{tokyo} that provides informations on all observables
at a time, measuring a quantum state is the only way to check a state
preparation within a (generally nonorthogonal) set.  In turn, the use
of homodyne tomography to test state-preparation becomes a way to
check the operation of a quantum device that prepares a chosen state
from a given one. It is now natural to ask if eventually it would be
possible to recover a complete information on the quantum device
itself, namely to reconstruct the detailed form of its Hamiltonian---
or, more generally, of its Liouvillian, as in reality the device is
always an open quantum system. Previous theoretical proposals to give
a complete characterization of quantum processes have been made in
Refs. \cite{cirac,chuang}. There, the methods are restricted to
systems with finite dimensional Hilbert space, and the method does not
lead to an explicit reconstruction of the Liouvillian. 
In this letter we show how this goal can be achieved in practice,
presenting a simple experimental setup for measuring the Liouvillian
of a phase-insensitive optical device, using currently available technology.

\begin{figure}[hbt]\vskip .3truecm\begin{center}
\epsfxsize=.8\hsize\leavevmode\epsffile{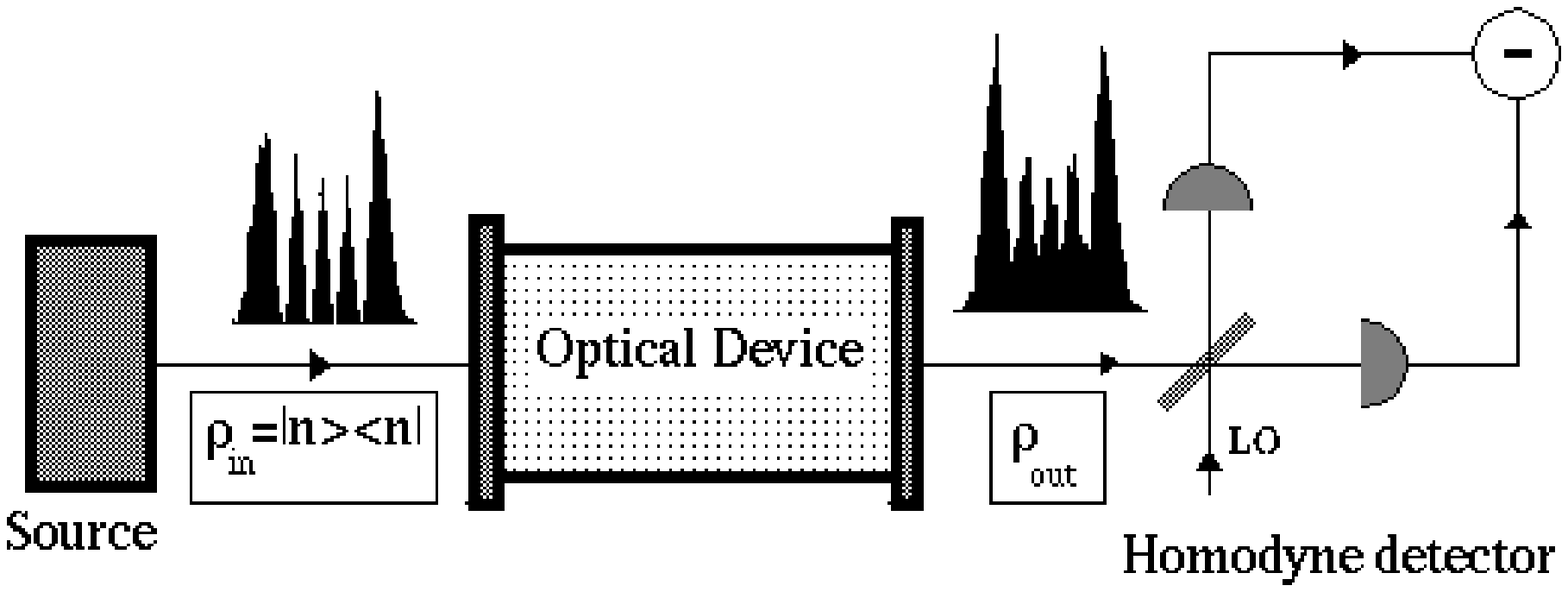}\end{center}
\begin{center}
\epsfxsize=.55\hsize\leavevmode\epsffile{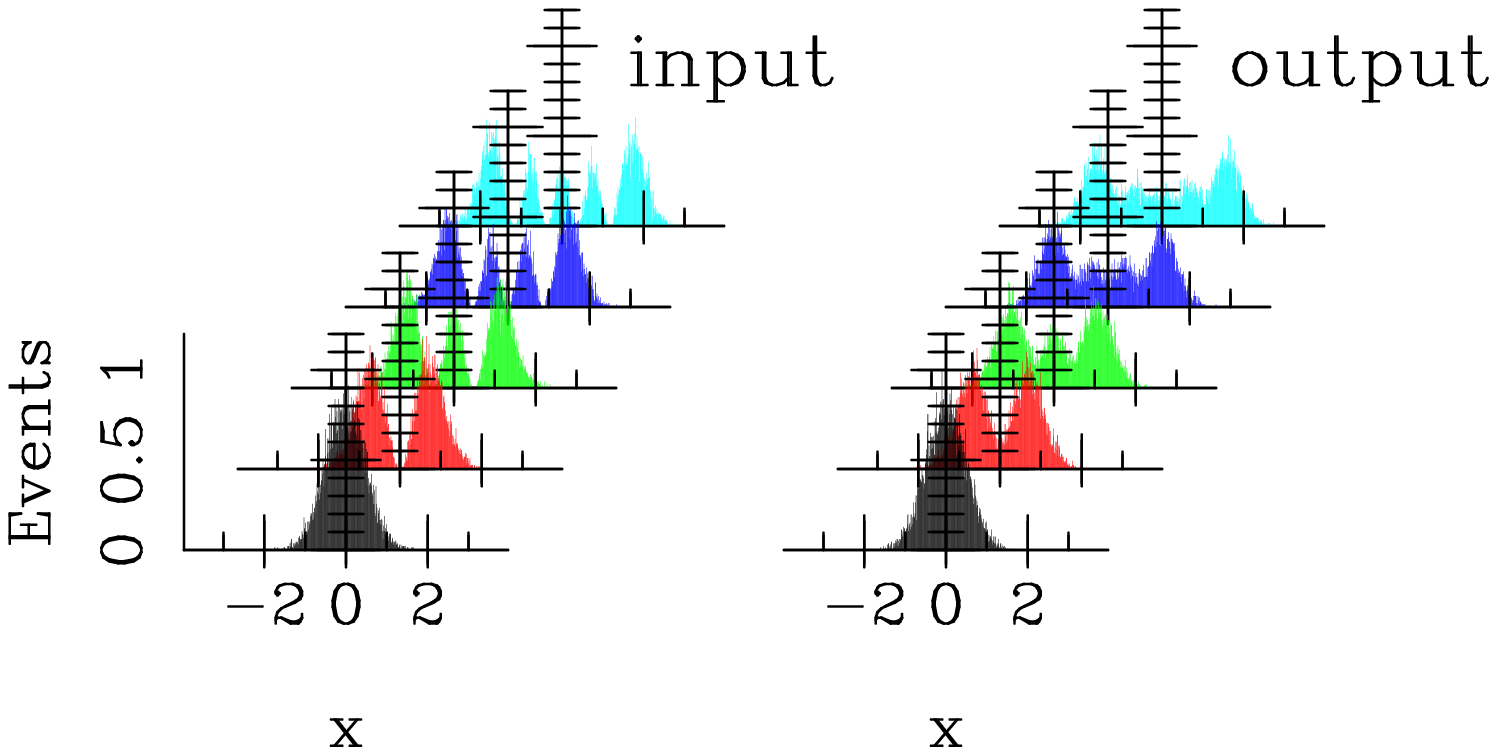}\end{center}
\caption{Sketch of the method for measuring the Liouvillian of an
optical device. A known input state $\rho_{in}$ is impinged into the
device, and a quantum tomography of the output state is performed
using a homodyne detector. By scanning an (over)complete set of states
$\rho_{in}$ at the input and comparing them with their respective
output states, it is possible to reconstruct the Liouvillian of the
device. The histograms of homodyne data, here given for the
sake of illustration, correspond to a device that consists of an empty 
cavity, and with the input states as number states
$\rho_{in}=|n\rangle\langle n|$.}\label{f:device}\end{figure}

The main idea for reconstructing the Liouvillian of a quantum device
is sketched in Fig. \ref{f:device}. One should impinge the device with
a known input state $\rho_{in}$ from a (over)complete set, then
determine the state $\rho_{out}$ at the output, and finally compare
$\rho_{in}$ to $\rho_{out}$. For an optical device the determination
of the output state is made possible by the homodyne-tomography
technique. Regarding the generation of the set of input states
$\{\rho_{in}\}$, an experimental method is suggested later 
in this letter. The evolution of the state from $\rho_{in}$ to $\rho_{out}$ is
governed by the Green (super)operator $\cal G$
\begin{eqnarray} \rho_{out}={\cal G}\rho_{in}\;,\label{evoluz}
\end{eqnarray} 
where $\cal G$ has actually a four-index matrix representation, and on
the Fock basis one has $\langle n|\rho_{out}|m\rangle =\sum_{h,k=0}^\infty\
{G}^{hk}_{nm}\langle h|\rho_{in}|k\rangle $.
For a device that is 
homogeneous along the direction of light propagation the Green
superoperator can be written as the exponential of a constant
Liouville superoperator $\cal L$ as follows
\begin{eqnarray} {\cal G}=\exp({\cal L}\tau) \;,\label{liouv}
\end{eqnarray} 
where $\tau$ is the propagation time ({\it i.e.} the device
length). The Liouvillian $\cal L$ gives the evolution of the state
through an infinitesimal slab of the device media according to the
master equation $\dot\rho={\cal L}\rho$. In this letter we restrict
our attention to the case of a perfectly phase-insensitive device: as
it will be clear from the following, the case of phase-sensitive
device is much more complicate, and will be analyzed elsewhere
\cite{maccone}. A phase-insensitive device, is a device that leaves
dephased states as dephased, as in the case of a traveling wave laser
amplifier. A dephased state is diagonal in the photon-number
representation, with density matrix of the form
$\rho=\sum_{n=0}^\infty\ r_n\ |n\rangle \langle n|$, where
$\{|n\rangle \}$ denotes the complete set of eigenvectors of the
photon-number operator $a^\dagger a$ of the field mode with
annihilation operator $a$. For the evolution of dephased states it is
sufficient to determine the sector of the Green superoperator that
evolves dephased states, {\it i.e.} the two-index Fock matrix
$G_{nm}=\langle n|\,\,{\cal G}[ \,|m\rangle\langle m|\,]\,\,|n\rangle$.
 \par
\begin{figure}[hbt]\vskip .3truecm\begin{center}
\epsfxsize=.84\hsize\leavevmode\epsffile{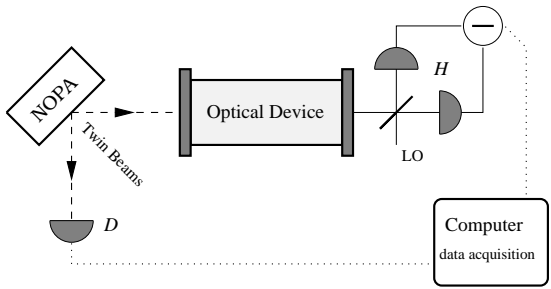}\end{center}
\caption{Experimental setup, including the apparatus used to generate
the input number states $\rho_{in}=|n\rangle \langle n|$ needed for
the tomographical reconstruction of the Liouvillian of an optical
device. A random-$n$ Fock state $|n\rangle$ for the input beam
is achieved by performing photodetection at $D$ on the other twin
beam, $n$ being the measured number of photons. A non degenerate
optical parametric amplifier (NOPA) with vacuum input 
is used to produce the twin beams. }
\label{f:fockgen}\end{figure}

 The experimental reconstruction of
$L_{nm}$ could be performed by impinging a number state
$\rho_{in}=|n\rangle\langle n|$ on the device, and then making the
homodyne tomography of the output state. In this fashion, the number
probability distribution $r_k(n)$ of the output coincides with the
$n$-th row of the Green matrix $G_{nm}$, and by varying $n$ one would
reconstruct the whole matrix. Since producing number states is
experimentally difficult, one would try to use coherent states
instead. In this way matrix elements of the form $\langle\psi |\,\,{\cal
G}[ \,|\alpha\rangle\langle\alpha|\,]\,\,|\psi'\rangle$ would be
obtained, with $|\alpha\rangle$ denoting the scanning coherent input
state, and $|\psi\rangle$ and $|\psi'\rangle$ being a couple of
vectors of the tomographically reconstructed matrix
representation. Unfortunately, the relation between $G_{nm}$ and
$\langle n|\,\,{\cal G}[\, |\alpha\rangle\langle\alpha|\,]\,\,|n\rangle$
is highly singular, involving the $P$-function of $|m\rangle\langle
m|$, and hence the matrix $G_{nm}$ cannot be obtained in this way
starting from experimental data. On the other hand, the Fock
representation has a privileged role, because here the Liouvillian
matrix has a transparent meaning in terms of creation and annihilation
operators.  How to overcome the problem of generating input
number-states?  Actually, for our purpose, it is sufficient to
generate number states with random $n$ as far as $n$ is known. This
leads us to devise the setup depicted in Fig. \ref{f:fockgen}.
A non degenerate optical parametric amplifier (NOPA) with a strong
classical pump down-converts the vacuum into a pair of twin beams. The
twin beams are used as a random-$n$ Fock state generator, by measuring
the number of photons on one beam (detector $D$ in Fig.
\ref{f:fockgen}) while impinging the other beam on the optical
device. For quantum efficiency $\eta_D=1$ at detector $D$, the
photodetection would reduce the twin-beam state $|\mbox{TB}\rangle
\propto \sum^\infty_{n=0}\kappa^n|n,n\rangle $ into a random-$n$ state
$|n\rangle $ at the input of the optical device, with thermal
probability distribution $w_n=|\kappa|^{2n}\left(1-|\kappa|^2\right)$,
where $n$ is the measurement outcome at $D$. The tomographically
reconstructed number probability $\langle k|\,\,{\cal
G}[\,|n\rangle\langle n|\,]\,\,|k\rangle$ of the output state already
would provide the $n$th row $G_{kn}$ of the Green matrix. On the other
hand, for $\eta_D<1$, a mixed state $\rho_n$ will actually enter the
device instead of $|n\rangle\langle n|$, as a result of state
reduction at $D$. The outcome $n$ probability distribution then
becomes
\begin{equation}
p_n=(1-|\kappa|^2)
\frac{(\eta_D|\kappa|^2)^n}{[(\eta_D-1)|\kappa|^2+1]^{n+1}}\;.
\end{equation}
One can easily show that the tomographically reconstructed output
number probability $r_k(n)\equiv\langle
k|{\cal G}[\rho_n]|k\rangle$ is related to the Green matrix through
the identity
\begin{eqnarray}
r_k(n)&=&[(\eta_D-1)|\kappa|^2+1]^{n+1}\nonumber\\
&\times&\sum_{m=0}^{\infty}
{m+n\choose n}[|\kappa|^2(1-\eta_D)]^m G_{k,m+n}\;.\label{GG}
\end{eqnarray}
The relation (\ref{GG}) can be inverted as follows
\begin{eqnarray}
G_{kl}&=&\frac{1}{[(\eta_D-1)|\kappa|^2+1]^{l+1}}\nonumber\\&\times&
\sum_{n=0}^{\infty}
{n+l\choose l}\left[\frac{(\eta_D-1)|\kappa|^2}{(\eta_D-1)|\kappa|^2+1}\right]^n
r_k(n+l)\;.\label{GGinv}
\end{eqnarray}
Eq. (\ref{GGinv}) is our algorithm to reconstruct the Green matrix
$G_{kl}$ from the collection of all tomographically measured number
probabilities $r_k(n)$ for different outcomes $n$ (in practice the sum
in Eq. (\ref{GGinv}) is truncated at some maximum $n$).  Notice the
interplay of the gain $\kappa$ and the quantum efficiency $\eta_D$ in
determining the probability $p_n$ on one hand, and in producing the
statistical errors in the reconstructed Green matrix on the other
hand. For decreasing quantum efficiency $\eta_D\to0$, larger values of
$n$ can be made more probable by increasing the gain of the NOPA as
$\kappa\to1^-$. However, at the same time, convergence of the sum in
Eq. (\ref{GGinv}) becomes slower, and statistical errors of matrix
elements $G_{kl}$ increase as result of tomographic errors on
$r_k(n)$. Hence, the effect of quantum efficiency $\eta_D$, which
reduces the size of the viewable matrix $G_{kl}$, can be partially
compensated by increasing the gain of the NOPA, however at expense of
statistical errors for $G_{kl}$. For the tomographic measurement, by
increasing the number of experimental data and using
$\eta_H$-dependent pattern functions, the method of
Ref. \cite{dariano3} can compensate the effect of low quantum
efficiency $\eta_H<1$, that, anyhow, must be above the threshold
$\eta_H=1/2$. On the other hand, for quantum efficiency $\eta_D$ at
detector $D$ there is not such a threshold, as one can see from
convergence and and error-propagation analysis of Eq. (\ref{GGinv}).
\par The proposed state-reduction scheme---based on twin-beams from a
NOPA---is not a new one, and, for example, a similar setup has been
proposed in Ref.\cite{caves} to generate Schr\"odinger-cat states. As
such state-reduction is the core of our measurement method, we want to
examine it at work in a realistic situation. Typically the NOPA can be
pumped by the second harmonic of a Q-switched mode-locked Nd:YAG
laser, with the output twin-beams pulsed at a repetition rate of
80 MHz, and with a 7ps pulse duration. Thus, the twin-beam mode with
annihilator $a$ at the input of the optical device is actually a
wideband mode, with frequency centered around 532nm, and width of
140 GHz (the inverse of the pulse time-length). The same Nd:YAG laser
beam is used for the local oscillator (LO) of the homodyne detector
$H$. In this way the LO has the same central frequency and the same
time-envelope of the beam at the input of the optical device. The
integration time at photodetector $D$ can be set to 1 ns, which is
greater than the pulse width and shorter than the distance between
pulses. In this way each pulse is completely annihilated by the
detector $D$ during the integration time, and, correspondingly, the
homodyne measurement is made with the LO matched on the same pulse
shape of the signal twin-beam, which means that the measurement is
performed on the right wideband mode after reduction by detector
$D$. Moreover, the detector $D$ and the optical device can be
geometrically placed in such a way that, within a narrow solid angle,
the direction of the respective input $k$-vectors are the same
relative to the $k$-vector of the NOPA pump, so that state reduction
at $D$ affects only radiation at the twin $k$ at the input of the
device, so that the state-reduced modes at $D$ and at the device are
perfectly matched.  \par From the above scenario it follows that the
mode with annihilator $a$ at the input of the optical device is
actually a wideband mode, and hence we measure the effective
Liouvillian over a 140 GHz bandwidth centered around 532nm. Then, it is
clear that all measurements for different random inputs can be
considered as independent only if the (atomic) relaxation times in the
optical devices are shorter than the pulse-repetition period.  \par
Now we show results from some Monte-Carlo simulated experiments to see
our method at work, and estimate the number of measurements needed for
the reconstruction of $G_{kl}$.  \par The simplest phase insensitive
device is the phase-insensitive linear amplifier (PIA), with
Liouvillian $\cal L$
\begin{eqnarray}
{\cal L}=2\left\{A{\cal D}[a^\dagger ]+B{\cal D}[a]\right\}
\;,\label{piadef}
\end{eqnarray} 
where ${\cal D}[\theta]\rho\doteq\theta\rho\theta^\dagger-
1/2(\theta^\dagger \theta\rho+\rho\theta^\dagger \theta)$ for any
complex operator $\theta$. The Liouvillian matrix has the form
\begin{eqnarray}
&L_{nm}\doteq\langle n|{\cal L}\Big[ |m\rangle\langle
m|\Big]|n\rangle = \label{piamatrix}&\\
&2\Big\{A(m+1)[\delta_{nm+1}-\delta_{nm}]+
Bm[\delta_{nm-1}-\delta_{nm}]\Big\}&\nonumber\;,
\end{eqnarray}
where $\delta_{ik}$ denotes the Kroneker delta. Notice that $L_{nm}$ is
tridiagonal, the upper diagonal corresponding to the one-photon
absorption $a\rho a^\dagger$ of the loss term ${\cal D}[a]$,
the lower diagonal corresponding to the one-photon 
emission $a^\dagger\rho a$ of the gain term ${\cal D}[a^\dagger]$, and
the main diagonal containing the anticommutators coming from both
terms.
\begin{figure}[hbt]\vskip .3truecm\begin{center}
\epsfxsize=.45\hsize\leavevmode\epsffile{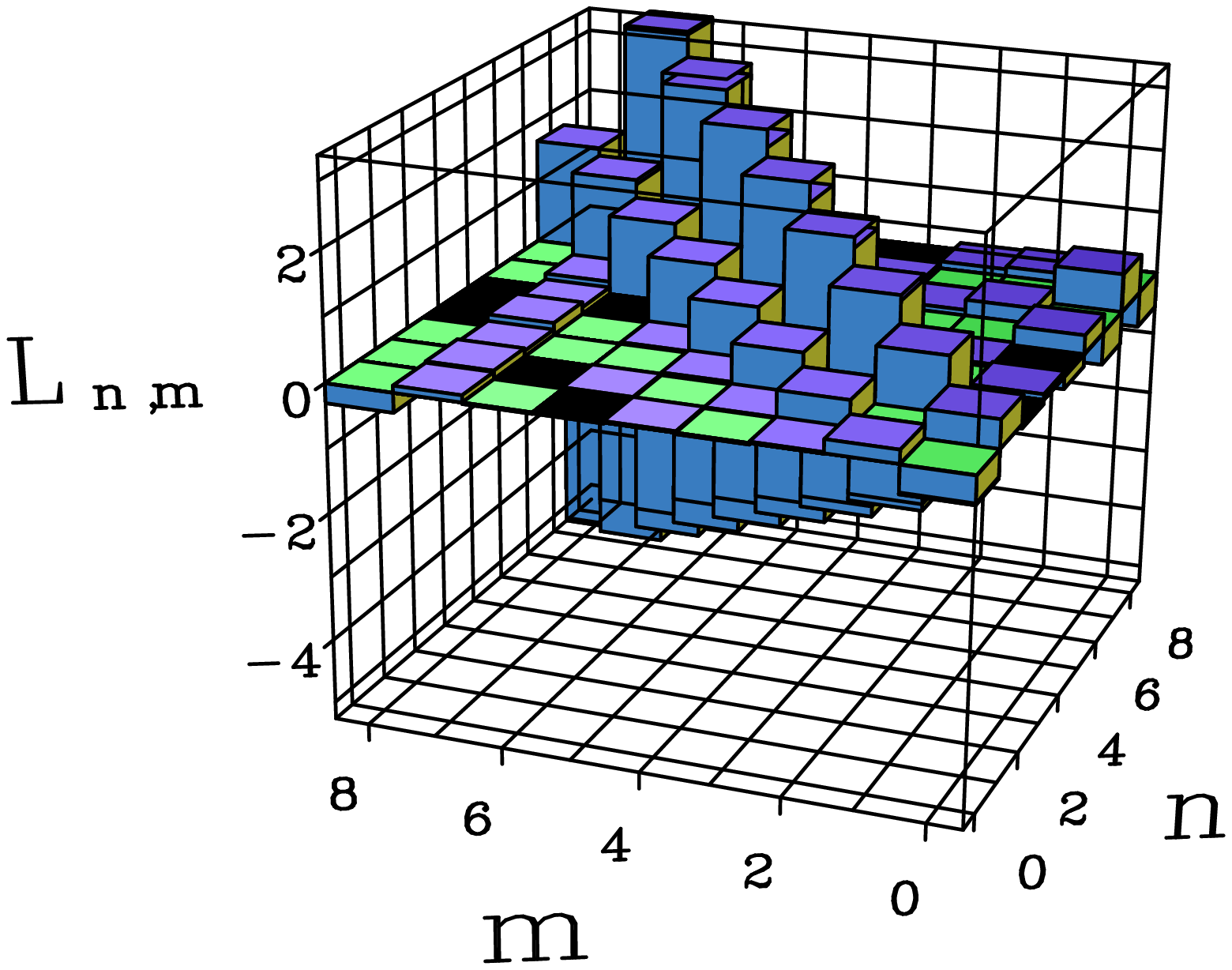}
\epsfxsize=.5\hsize\leavevmode\epsffile{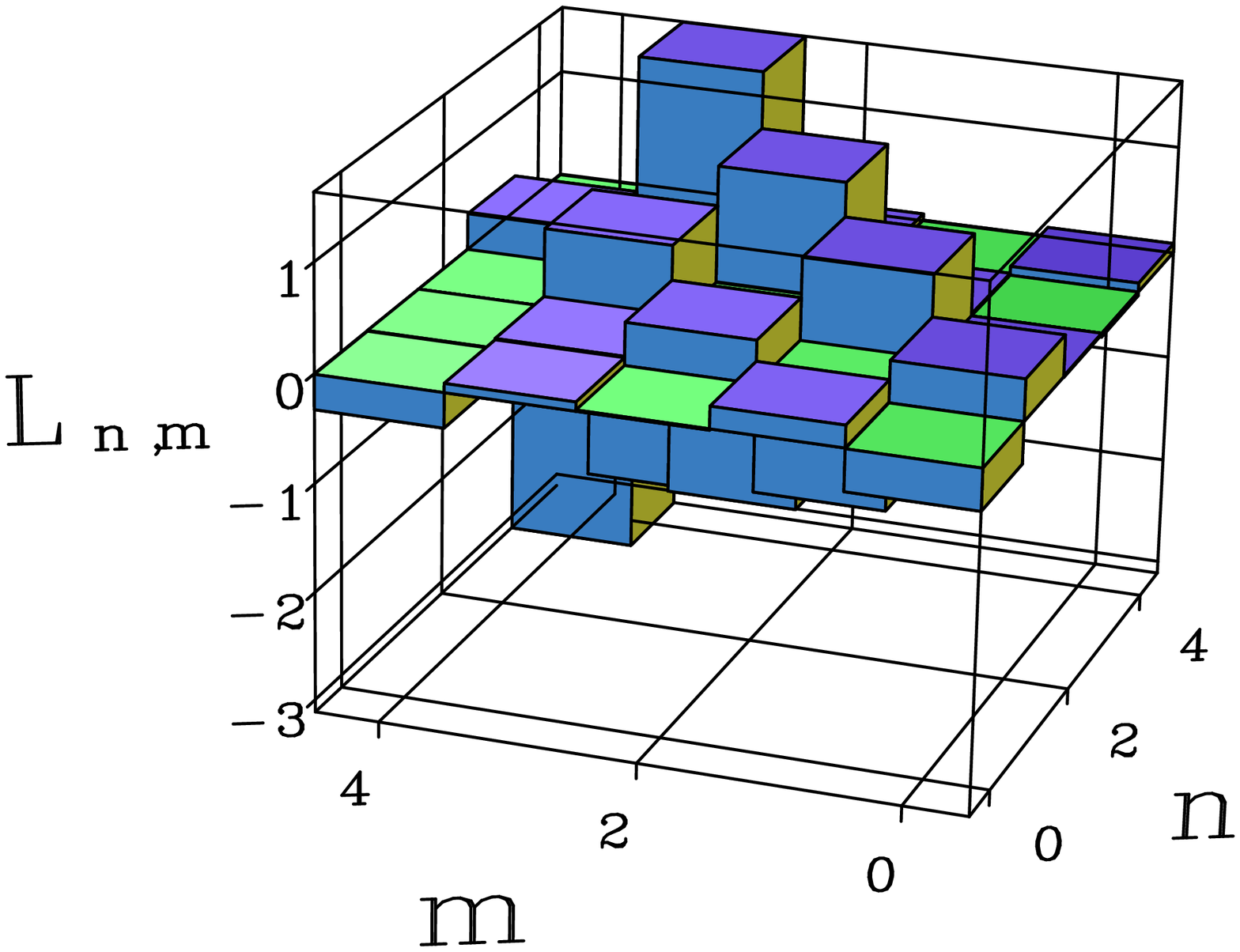}
\vskip -.7truecm\hskip -3.5truecm (a)\hskip 3.5truecm (b)
\end{center}
\vskip .5truecm
\caption{(a) Linear phase insensitive laser amplifier with $A=.1940$
and $B=.00945$. Monte Carlo simulation of the reconstruction of the
Liouvillian $L_{nm}$ in Eq. \ref{piamatrix} by means of the proposed
experimental setup. The reconstruction is performed by using 2
statistical blocks of $10^6$ homodyne data for each of the output
states (from a total of $2.7*10^{11}$ of data). Quantum efficiencies
are $\eta_D=.8$ and $\eta_H=.85$. The gain of the NOPA is
$\kappa=.6$. (b) the same reconstruction for lower quantum efficiency
$\eta_D=.3$. Here the gain is set to $\kappa=.4$, and the same number
of data is used out of a total of $9.9*10^{11}$. }
\label{f:tomopia}\end{figure}
\begin{figure}[hbt]\vskip .3truecm\begin{center}
\epsfxsize=.5\hsize\leavevmode\epsffile{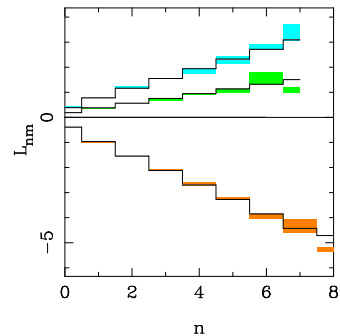}
\end{center}
\caption{The three main diagonals of the Liouvillian $L_{nm}$ of
Fig. \protect{\ref{f:tomopia}a} are given with their statistical error
bars. The full line is the theoretical value from
Eq. (\ref{piamatrix}).}
\label{f:tomotridiag}\end{figure} 
In Fig \ref{f:tomopia}a we show a typical result of a Monte Carlo
experiment of the tomographic reconstruction of the Liouvillian
(\ref{piamatrix}). We used quantum efficiency $\eta_D=.8$ and
$\eta_D=.3$ at detector $D$ and $\eta_H=.85$ at the homodyne detector
$H$. One can see that the details of the matrix are well recovered,
and the truncation of the Hilbert space dimension does not affect the
reconstruction. In Fig. \ref{f:tomotridiag} the three main diagonals
of the matrix are plotted with their statistical errors against the
theoretical value, showing a very good agreement. The statistical
errors are of the same size of those of the tomographically
reconstructed output probabilities. Notice that the number of data
$10^{11}\div 10^{12}$ needed for this experiment could be collected in
a few minutes at the repetition rate of 80 MHz.  In
Fig. \ref{f:tomopia}b we present a simulated experiment using the very
realistic value $\eta_D=.3$ of quantum efficiency, however for the
reconstruction of a smaller matrix $5\times 5$. Notice that only the
photodetector $D$ is required to be linear single-photon resolving,
whereas the homodyne detector takes advantage of amplification from
the LO, and hence can use high efficiency detectors (the value
$\eta_H=.85$ here used has been widely surpassed in the real
tomographic experiments, as in Ref. \cite{schiller}).  

\par As another example, we simulated the experimental tomography of
the effective Liouvillian of a one-atom traveling-wave laser
amplifier.
\begin{figure}[hbt]\vskip .3truecm\begin{center}
\epsfxsize=.6\hsize\leavevmode\epsffile{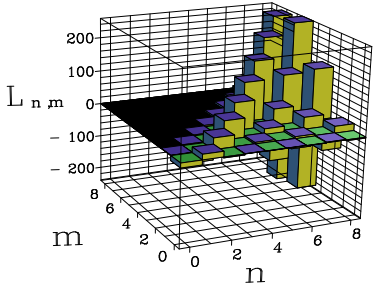}
\end{center}
\caption{Theoretical Liouvillian for a one-atom laser, obtained by
solving numerically the master equation (\ref{lasermeq}). The
parameters for this laser are $C\doteq\frac{g^2}{\gamma\gamma_\perp}=12$;
$n_s\doteq\frac{\gamma_\parallel\gamma_\perp}{4g^2}=7$;
$\sigma_0=1;\; f=\frac{\gamma_\parallel}{2\gamma_\perp}=1$;
$\gamma=1$; $t_*=.0115$.}\label{f:liouvtheor}\end{figure}
\begin{figure}[hbt]\vskip .3truecm\begin{center}
\epsfxsize=.6\hsize\leavevmode\epsffile{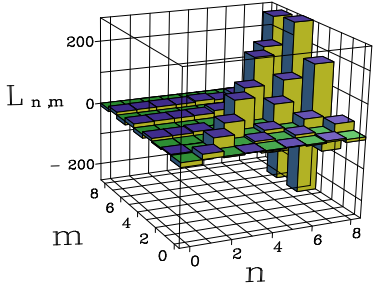}\end{center}
\caption{Monte Carlo simulated experiment for the reconstruction of
the laser theoretical Liouvillian in Fig. \protect{\ref{f:liouvtheor}}. 
Here $\eta_D=.8$, $\eta_H=.85$, and $\kappa=.65$. A set of $8*10^7$
homodyne data have been used of a total of $3.7*10^{11}$ measurements
with random photon number at detector $D$.}
\label{f:liouvtomo}\end{figure}
In Fig. \ref{f:liouvtheor} the theoretical Liouvillian matrix is
plotted, as obtained from a long-run quantum jump
simulation of the one-atom-laser master equation
\cite{englert}:
\begin{eqnarray}
&\dot\rho=\bigg\{\frac{\gamma_\|}2(1+\sigma_0){\cal D}[\sigma_+]+
\frac{\gamma_\|}2(1-\sigma_0){\cal D}[\sigma_-]+&\nonumber \\
&\frac 14\left(\gamma_\perp-\frac {\gamma_\|}2\right){\cal D}[\sigma_z]
+\gamma{\cal D}[a]\bigg\}\rho+g[\sigma_+a-\sigma_-a^\dagger,\rho]&
\;\label{lasermeq}
\end{eqnarray}
where $g$ is the electrical-dipole coupling, $\gamma_\|$ and
$\gamma_\perp$ are the decay rates of population inversion and atomic
polarization respectively, $\gamma$ is the cavity decay rate,
$\sigma_0$ is the unsaturated inversion ($-1\leq\sigma_0\leq 1$),
$\sigma_{\pm z}$ are the Pauli matrices (with $0, \pm 1$ entries), and
$\rho$ now denotes the joint atom-radiation density matrix. In the
quantum jump simulation, the atom is traced out at a time
$t_*\gg\gamma^{-1}_{\|,\perp}$.  In Fig. \ref{f:liouvtomo} a Monte
Carlo simulated tomographic experiment is shown for reconstructing the
Liouvillian in Fig. \ref{f:liouvtheor} (the output homodyne
probabilities are simulated starting from the quantum jump Green
matrix). One can see how the method allows a detailed reconstruction
of $L_{nm}$, including not only one-photon processes on the three main
diagonals, but also multiphoton-absorptions on the upper triangular
part.  \par In conclusion, we have seen that it is possible to
experimentally reconstruct the Liouvillian of a quantum optical
phase-insensitive device, using homodyne tomography in a scheme based
on parametric down conversion from a NOPA. We have shown the
feasibility of the reconstruction with an experimental setup that uses
standard technology devices. The problem of low efficiency at the
single-photon resolving detector $D$---the major obstacle for the
experiment---has been solved by implementing a compensation algorithm
that makes the reconstruction of $5\times 5$ Liouville matrix possible
even for $\eta_D=.3$, and with a number of data that can be collected
in a few minutes of experimental run.

\end{document}